\documentstyle[12pt]{article}
\begin{document}
\tolerance=5000
\def\be{\begin{equation}}
\def\ee{\end{equation}}
\def\bea{\begin{eqnarray}}
\def\eea{\end{eqnarray}}
\def\nn{\nonumber \\}
\def\cF{{\cal F}}
\def\det{{\rm det\,}}
\def\Tr{{\rm Tr\,}}
\def\e{{\rm e}}
\def\str{{\rm STr}}
\def\diag{{\rm diag}}
\def\A{{\cal A}}
\def\rag{\sqrt{R^4 + {\A^4 \over 4g^2}}}
\def\ttta{{g R^2 \over \A^2}
\ln\left\{{1 \over R^2}\left({\A^2 \over 2g } + \rag\right) \right\}}
\def\tttb{{g R^2 \over 2\A^2}
\ln\left\{{1 \over R^2}\left({\A^2 \over 2g } + \rag\right) 
\right\}}
\def\tttc{{g \over 2\A^2}
\ln\left\{{1 \over R^2}\left({\A^2 \over 2g } + \rag\right) 
\right\}}
\def\ttrag{\left[{1 \over 4}\rag + {g R^4 \over 2 \A^2}
\ln\left\{{1 \over R^2}\left({\A^2 \over 2g} + \rag\right) 
\right\}\right]}

\ \hfill
\begin{minipage}{3.5cm}
NDA-FP-34 \\
June 1997 \\
\end{minipage}

\vfill

\begin{center}
{\Large\bf On the instability of effective potential \\
for non-abelian toroidal D-brane}

\vfill

{\large\sc Shin'ichi NOJIRI}\footnote{e-mail : nojiri@cc.nda.ac.jp} 
and {\large Sergei D. ODINTSOV$^\clubsuit$}\footnote{e-mail : 
odintsov@quantum.univalle.edu.co, 
odintsov@kakuri2-pc.phys.sci.hiroshima-u.ac.jp}

\vfill

{\large\sl Department of Mathematics and Physics \\
National Defence Academy \\
Hashirimizu Yokosuka 239, JAPAN}

{\large\sl $\clubsuit$ Tomsk Pedagogical University \\
634041 Tomsk, RUSSIA \\
and \\
Dep.de Fisica \\
Universidad del Valle \\
AA25360, Cali, COLOMBIA \\
}

\vfill

{\bf abstract}

\end{center}

We calculate the one-loop effective potential for the toroidal non-abelian D-brane
in the constant magnetic SU(2) gauge field.
The study of its properties shows that the potential is 
unbounded below. 
This fact indicates the instability of the non-abelian 
D-brane in the background under consideration like the 
instability of chromomagnetic vacuum in SU(2) gauge theory.

\newpage

One of the key ingredients in the study of the 
dynamics of $D=11$ M-theory \cite{1} is 
D-brane \cite{2,3,4} which represents the 
solitonic solution of string theory.
The effective theories of D-branes have been the 
subject of much recent study (see \cite{5} and 
references therein).

The study of the effective potential in theory of extended 
objects in one-loop or large $d$-approximation may 
clarify the quantum properties of these objects.
For string case, large $d$ approximation has been 
developed in ref.\cite{7} as systematic expansion 
for the effective action in powers of $1/d$. 
The static potential may be then obtained by studying 
the saddle point equations for the leading order term.
The similar program may be realized for membranes \cite{8,9}, 
$p$-branes \cite{9} and abelian D-branes \cite{10}.

The non-abelian generalization of DBI-theory has been 
recently presented in ref.\cite{6}. 
That is the purpose of the present note to look to the 
properties of the effective potential in the non-abelian 
toroidal D-brane in the constant magnetic SU(2) field.
We find that the correspondent potential shows the 
instability in the background under discussion.

The action which corresponds to the non-abelian 
generalization of 
Born-Infeld D-brane is given by \cite{6}
\bea
\label{nDaction}
S&=&k \int d\zeta^{p+1} \nn
&& \times
\str \Bigl[\sqrt{\det \left( G_{ij}
+D_i X^m\left(\delta^{mn} - {i \over g} 
[X^m, X^n]\right)^{-1}
 D_j X^n + {F_{ij} \over g} \right)} \nn
&& \hskip 2cm \times\sqrt{\det \left(\delta^{mn} 
- {i \over g} [X^m, X^n]\right)}\Bigr]\ .
\eea
Here we consider the action in the Euclidean signature 
and $m,n=p+1,p+2,\cdots,d-1$ ($d$ is 
the space-time dimension), $i,j=0,1,2,\cdots,p$. 
$G_{ij}$ is the metric on the D-brane world volume, 
$X^m$ belong to the adjoint representation of the 
non-abelian algebra, $F_{ij}$ is the
field strength of the algebra
and $g$ is the coupling constant.
$\det$ is the determinant with respect to the indeces 
$i$, $j$.
$\str$ is the trace about the matrix which is 
the representation of the algebra,  
after all it should be symmetrized with respect to 
$F_{ij}$, $D_i X^m$ and $[X^m, X^n]$.

In this letter, we concentrate, for simplicity, 
on case that $p=2$ (membrane) and the 
non-abelian algebra is SU(2).
We also assume the metric $G_{ij}$ 
is given by $G_{ij}=R_i R_j \delta_{ij}$ (here the 
sum convention is not used and $R_i$'s 
are constants and we normalize $R_0=1$).  The
coordinates satisfy the periodic boundary 
condition (toroidal D-brane):
\bea
\label{pb}
&& X^m(\zeta^0, \zeta^1, \zeta^2, \cdots, \zeta^p)
 =X^m(\zeta^0+T, \zeta^1, \zeta^2, \cdots, \zeta^p) \nn
&& =X^m(\zeta^0, \zeta^1+1, \zeta^2, \cdots, \zeta^p) 
 =X^m(\zeta^0, \zeta^1, \zeta^2+1, \cdots, \zeta^p) \nn
&& = \cdots =X^m(\zeta^0, \zeta^1, \zeta^2, \cdots, \zeta^p+1) \ .
\eea
Therefore $R_i$ $(i\neq 0$) is the period with respect to the 
world volume coordinate $\zeta^i$.
We consider the one-loop 
effective potential 
in the non-abelian constant magnetic background:
\bea
\label{magbg}
&& A^B_1={1 \over 2}\A \sigma^1\ ,\ \ 
A^B_2={1 \over 2}\A \sigma^2\ ,\nn 
&& F^B_{ij}={1 \over 2} f^B_{ij} \sigma^3\ ,\ \ 
f^B_{12}=-i[A_1, A_2]={1 \over 2}\A^2 \sigma^3 \nn
&& X^{B\,m}=0 \ .
\eea
Here $\sigma^a$'s ($a=1,2,3$) are Pauli matrices. 
in the following, we abbreviate the index $m$ of $X^m$ 
for the simplicity in the notation.

The effective potential is defined by \cite{7}
\be
\label{EP}
V=-\lim_{T\rightarrow\infty}{1 \over T}
\ln \int {\rm D}X\e^{-S}\ .
\ee
Here we treat the gauge field as a classical external 
field. The quantum contribution from the gauge fields will be 
taken into account below.

Let decompose $D_i X$ in the components
\bea
\label{dcp}
D_i X &\equiv& \partial_i X - i [A_i, X] \nn
&=& {1 \over 2}D_i X^a \sigma^a \nn
&=& {1 \over 2}(\partial_i X^a 
+ \epsilon^{abc} A^b_i X^c ) \sigma^a\ .
\eea
We expand the gauge field and the field strength
arround the non-abelian magnetic background 
(\ref{magbg})  as follows
\bea
\label{exp}
&& A_1\rightarrow {1 \over 2}\A \sigma^1
+{1 \over 2}A^a_1\sigma^a\ ,\ \ 
A_2={1 \over 2}\A \sigma^2 +{1 \over 2}A^a_2\sigma^a
\ ,\nn 
&& F_{0i}={1 \over 2}\partial_0 A^a_i \sigma^a\ ,\ \ 
F_{12}={1 \over 2}\A^2 \sigma^3 + {1 \over 2}
F^a_2\sigma^a \ .
\eea
Here we have chosen the $A_0=0$ gauge and
\bea
\label{fstr}
F^1_{12}&=& \partial_1 A^1_2 - \partial_2 A^1_1 
+ A^2_1A^3_2 - A^3_1 (\A + A_2^2) \nn
F^2_{12}&=& \partial_1 A^2_2 - \partial_2 A^2_1 
+ A^3_1A^1_2 - (\A + A_1^1)A^3_2  \nn
F^3_{12}&=& \partial_1 A^3_2 - \partial_2 A^3_1 
+ \A(A^1_1 + A^2_2 ) + A^1_1A^2_2 - A_1^2A^1_2  \ .
\eea

First we should note
\bea
\label{str}
\str \left(D_i X D_j X (F^B_{12})^{2n}\right)
&=&{1 \over 2}\left({\A^2 \over 2}\right)^{2n}D_i X^3 D_j X^3
\nn &&+{1 \over 2(2n+1)}
\left({\A^2 \over 2}\right)^{2n}\sum_{a=1,2}
D_i X^a D_j X^a \nn 
\str \left(F_{ij}F_{kl} (F^B_{12})^{2n}\right)
&=&{1 \over 2}\left({\A \over 2}\right)^{2n}F_{ij}^3 
F_{kl}^3 \nn
&& +{1 \over 2(2n+1)}\left({\A \over 2}\right)^{2n}
\sum_{a=1,2}F_{ij}^a F_{kl}^a\ .
\eea
Therefore if $f(x)$ is an arbitrary even 
($f(-x)=f(x)$) function, we find
\bea
\label{inteq}
\str \left(D_i X D_j X f(F^B_{12})\right)
&=&{1 \over 2}f\left({\A^2 \over 2}\right)D_i X^3 D_j X^3 \nn
&& +{1 \over A^2}\int_0^{\A^2 \over 2} f(x)dx 
\sum_{a=1,2}D_i X^a D_j X^a \nn 
\str \left(F_{ij}F_{kl} f(F^B_{12})\right)
&=&{1 \over 2}f\left({\A \over 2}\right)F_{ij}^3 
F_{kl}^3 \nn
&& +{1 \over A^2}\int_0^{\A^2 \over 2} f(x)dx 
\sum_{a=1,2}F_{ij}^a F_{kl}^a\ .
\eea

Then expanding the action (\ref{nDaction}) 
in the background (\ref{magbg})
up to the terms quadratic with respect to the fields, 
we obtain
\bea
\label{det}
S&=&k \int d\zeta^3 
\str \sqrt{\det \left( G_{ij}
+D_i X D_j X + F_{ij} \right)} \nn
&=&k \int d\zeta^3 
\left[2\sqrt{R^4+{\A^2 \over 4g^2}} + L^X (X^m) 
+ L^A (A_i) \right]  +\ \cdots 
\eea
\bea
\label{LagX}
&& L^X (X^m) \nn
&& = {1 \over 2}\rag 
(D_0 X^3)^2 + {R^2 \over 2\rag}
\{(D_1 X^3)^2 + (D_2 X^3)^2\} \nn
&& +\sum_{a=1,2}\left[ \ttrag (D_0 X^a)^2 \right. \nn
&& \left. + \ttta \{(D_1 X^a)^2 + (D_2 X^a)^2\} \right] \nn
&& \equiv \hat A 
(D_0 X^3)^2 + \hat B\{(D_1 X^3)^2 + (D_2 X^3)^2\} \nn
&& + \sum_{a=1,2}[\hat C
(D_0 X^a)^2 + \hat D \{(D_1 X^a)^2 + (D_2 X^a)^2\} ] \\
\label{LagA}
&& L^A(A_i) \nn
&&={R^2 \over 4\rag}
\left( (F^3_{01})^2 + F^3_{02})^2 \right) 
+{1 \over 4\rag} (F^3_{12})^2 \nn
&&+ \tttb \sum_{a=1,2}\left( (F^a_{01})^2 
+ F^a_{02})^2 \right) \nn
&& + \tttc \sum_{a=1,2}(F^a_{12})^2
\nn
&& \equiv A^2 \left( (F^3_{01})^2 + F^3_{02})^2 \right) 
+ B^2 (F^3_{12})^2 \nn
&& + \sum_{a=1,2}\left\{
C^2 \left( (F^a_{01})^2 + F^a_{02})^2 \right) + D^2 
(F^a_{12})^2 \right\}\ . 
\eea
Here we assume $R_i=R$ ($i\neq 0$) and 
$\cdots$ denotes the higher order terms with 
respect to $X^m$, $A_i$ and total derivative terms.

We now consider the contribution from $X^m$ to 
the one-loop effective potential.
Eq. (\ref{LagX}) can be rewritten as follows
\bea
\label{LagX2}
&& L^X(X^m) \nn
&&= \hat A \left[(\partial_0 X^3)^2 
+ {\hat B \over \hat A}
\{(\partial_1 X^3)^2 + (\partial_2 X^3)^2\} 
+ {2 \A^2 \hat D  \over \hat A} (X^3)^2 \right] \nn
&& +\hat C \sum_{a=1,2}\left[ 
(\partial_0 X^a)^2 
+ {\hat D \over \hat C} 
\{(\partial_1 X^a)^2 + (\partial_2 X^a)^2\} 
+ { \A^2 \hat B \over \hat C}(X^a)^2 
\right] \nn
&& + (\hat B + \hat D)\A (X^2 
\partial_1 X^3 - X^3 \partial_1 X^2
- X^1 \partial_2 X^3 + X^3 \partial_2 X^1) \nn
&& + \mbox{total derivative terms} 
\eea
By diagonalizing (\ref{LagX2}) with respect to $X^a$, 
we find the following expression for the effective 
potential;
\bea
\label{VeffX}
V_X&=& V_X^0 + V_X^+ + V_X^- \\
\label{V0X}
V_X^0 &=& {(d-3)2\pi \over 2}\sqrt{\hat D \over 
\hat C}\sum_{n_1,n_2=-\infty}^\infty \left(
n_1^2+n_2^2 + {\A^2 \hat B \over \hat D}
\right)^{1 \over 2} \nn
&=& {(d-3)2\pi \over 2}\sqrt{\hat D \over 
\hat C} E\left({\A^2 \hat B \over \hat D} \right) \\
\label{VpmX}
V_X^\pm&=& {(d-3)2\pi \over 2}
\sum_{n_1,n_2=-\infty}^\infty \left[
\left({\hat B \over \hat A}+{\hat D \over \hat C}\right)
(n_1^2+n_2^2)
+\A^2 \left( {2\hat D \over \hat A} + {\hat B \over \hat C}
\right) \right. \nn
&& \pm \left\{ \left({\hat B \over \hat A}-{\hat D \over \hat C}\right)^2 (n_1^2+n_2^2)^2 \right. \nn
&& + \left({2\A^2 \hat B \hat D (2\hat C^2 + \hat A^2) \over 
\hat A^2 \hat C^2} 
+ {-8\hat B \hat D + 2 \hat B^2 \over \hat A \hat C} \right)
(n_1^2+n_2^2) \nn
&& 
\left. \left. + {\A^4 (2\hat D \hat C - \hat A \hat B)^2 \over 
\hat A^2 \hat C^2 } \right\}^{1 \over 2} \right]^{1 \over 2}
\eea
In (\ref{V0X}), 
we have used zeta function regularization (see \cite{11} 
for an introduction) and $E(q)$ is defined by
\bea
\label{Epstein2}
E(q)&\equiv&\sum_{n_1, n_2=1}^\infty 
\left\{n_1^2+n_2^2+ q\right\}^{1 \over 2}\nn
&=&
-q^{1 \over 2}-{\pi \over 3}q^{3 \over 2} \nn
&&-{8 \over \pi}\Bigl[
q^{1 \over 2}\sum_{k=1}^\infty k^{-1}
K_1\left(2\pi k\sqrt{q}\right) 
+q^{3 \over 4}\sum_{k=1}^\infty k^{-{3 \over 2}}
K_{-{3 \over 2}}\left(2\pi k\sqrt{q}\right) \nn
&&+2\sum_{k=1}^\infty k^{-1} \sum_{d|k}d^2 
\left(1+{q \over d^2}\right)^{1 \over 2}
K_1\left(2\pi k\sqrt{1 +{q \over d^2}}\right)\Bigr]\ .
\eea
Since $K_\nu (z)\sim \e^{-z}$ when 
$|z|\rightarrow \infty$, we find 
\be
\label{epinf}
E(q)\rightarrow -{\pi \over 3}q^{3 \over 2}
\ee
when $q\rightarrow +\infty$ and 
\be
\label{ep0}
E(0)={f_T(1,1) \over 2\pi}
\ee
Here 
\be
\label{fT}
f_T(1,1)=2\pi\sum_{n_1,n_2=-\infty}^\infty \left(
n_1^2+n_2^2 \right)^{1 \over 2}=-1.438\cdots\ .
\ee
Note that $E(q)$ is monotonically decreasing function 
therefore always negative when $q\geq 0$.
Note that, in the limit $\A\rightarrow 0$, 
Eq.(\ref{VeffX}) reproduces the result for $p$-brane in
\cite{12}, up to a factor 3 which corresponds to  dimension of the 
SU(2) group:
\be
\label{membrane}
V_X^0 = {(d-3)2\pi \over 2}\cdot 3 \cdot {f_T(0,0) \over R}
\ .
\ee

In order to calculate the contribution from the quantum 
gauge field, first we solve the constraint 
$\partial_i A_i^a=0$ obtained from 
the $A_0=0$ gauge condition non-locally 
(locally in the momentum space)
by introducing a scalar field $A^a$
\be
\label{cons}
(A^a_1, A^a_2)={1 \over \sqrt{-\Delta}}
(\partial_2 A^a, -\partial_1 A^a)\ \ 
(\Delta=\partial_1^2 + \partial_2^2)
\ee
and further redefining $A^0$, $A^l$ and $A^t$ as
\be
\label{Alt}
A^0 \equiv A A^3\ ,\ \  
A^l\equiv {C \over \sqrt{-\Delta}}
(\partial_1 A^1 + \partial_2 A_2) \ ,\ \ 
A^t\equiv {1 \over \sqrt{-\Delta}}
(\partial_2 A^1 - \partial_1 A_2) 
\ee
Then we can rewrite Eq.(\ref{LagA}) up to total derivative 
terms as follows
\bea
\label{LagA2}
&& L^A \nn
&&=(\partial_0 A^0)^2 + (\partial_0 A^t)^2 + 
(\partial_0 A^l)^2 \nn
&& +\left({B^2 \over A^2}(-\Delta) 
+ {D^2\A^2 \over A^2}\right)(A^0)^2 
+ {2\A(B^2+D^2) \over AC}A^0 \sqrt{-\Delta} A^t \nn
&& +\left({D^2 \over C^2}(-\Delta)+
{\A^2 B^2 \over C^2}\right)(A^t)^2 
+ {D^2 \over C^2}(-\Delta)(A^l)^2
\eea
The appearence of mass-like term for gauge field is seen 
in above equation. That is contrary to the abelian 
case \cite{10} 
where such term does not appear.
After diagonalizing (\ref{LagA2}) 
with respect to $A^0$ and $A^t$, we find the 
contribution from the gauge field to the one-loop 
potential is given by
\bea
\label{potg}
V_A&=&V_A^0 + V_A^+ + V_A^- \nn
V_A^0&=& {D \over C}f_T(0,0) \nn
V_A^\pm&\equiv&{1 \over \sqrt2}
\sum_{n_1, n_2=-\infty}^\infty 
\left[\left({B^2 \over A^2} 
+ {D^2 \over C^2}\right)(n_1^2+n_2^2)
+ \left({D^2 \over A^2} 
+ {B^2 \over C^2}\right)\A^2 \right. \nn
&& \pm\left\{\left({B^2 \over A^2} 
- {D^2 \over C^2}\right)^2(n_1^2+n_2^2)^2 \right. \nn
&& +2\left({B^2 D^2 \over A^4} 
+ {B^2 D^2 \over C^4}
+{B^4 + D^4 + 4 B^2 D^2 \over A^2 C^2}\right)
\A^2(n_1^2+n_2^2) \nn
&& \left. \left. + \left({D^2 \over A^2} 
- {B^2 \over C^2}\right)^2\A^4\right\}^{1 \over 2}
\right]^{1 \over 2}\ .
\eea

We now consider the asymptotic behavior of the 
one-loop potential.
First we should note that 
\bea
\label{asRinf}
&&\hat A,\ \hat C \rightarrow {R^2 \over 2},\ \ 
\hat B,\ \hat D \rightarrow {1 \over 2} \nn
&& A^2,\ C^2 \rightarrow {1 \over 4},\ \ 
B^2, D^2 \rightarrow {1 \over 4R^2}
\eea
when $R\rightarrow \infty$ and 
\bea
\label{asR0}
&& \hat A \rightarrow {\A^2 \over 4g}, \ \ 
\hat B \rightarrow {gR^2 \over \A^2}, \ \ 
\hat C \rightarrow {A^2 \over 8g}, \ \ 
\hat D \rightarrow {gR^2 \over \A^2}\ln {\A^2 \over gR^2} \nn
&& A^2\rightarrow {gR^2 \over 2\A^2}, \ \ 
B^2 \rightarrow {g \over 2\A^2} \nn 
&& 
C^2 \rightarrow {gR^2 \over 2\A^2}\ln {\A^2 \over gR^2}, \ \ 
D^2 \rightarrow {g \over 2\A^2}\ln {\A^2 \over gR^2}\ .
\eea
Therefore the asymptotic behavior of the potential is given 
by 
\bea 
\label{asVinf}
&& 
V_X^0 \rightarrow {2\pi(d-3) \over 2} 
\cdot {E(\A^2)  \over R} \nn 
&& V_X^+ \rightarrow {2\pi(d-3) \over 2} \cdot 
{E(2\A^2) \over R} , 
\ \ 
V_X^- \rightarrow {2\pi(d-3) \over 2} \cdot 
{E(\A^2) \over R}  \nn
&& V_A^0 \rightarrow {f_T(1,1) \over R},\ \ 
V_A^\pm \rightarrow {2\pi E(\A^2) \over R}
\eea
when $R\rightarrow \infty$ and 
\bea 
\label{asV0}
V_X^0 &\rightarrow& {2\pi(d-3) \over 2} \cdot 
{2\sqrt 2 gR \over \A^2}\ln {\A^2 \over g R^2} 
E\left({\A^2 \over \ln {A^2 \over gR^2} }\right) \nn
&& \sim  -{2\pi(d-3) \over 2} 
{2\sqrt 2 \pi gR \over 3\A^2}\ln {\A^2 \over g R^2} 
\left({\A^2 \over \ln {A^2 \over gR^2} }\right)^{3 \over 2}
\nn 
V_X^\pm &\rightarrow& {2\pi(d-3) \over 2} \cdot 
{2g R \over \A^2}\ln {\A^2 \over g R^2} \nn
&& \times \sum_{n_1, n_2=-\infty}^\infty \left\{ n_1^2+n_2^2 
+ \A^2 \pm \left( \left(
n_1^2+n_2^2 \right)^2 + \A^4 \right)^{1 \over 2} 
\right\}^{1 \over 2} \nn
V_A^0 &\rightarrow& {1 \over R}f_T(0,0) \nn
V_A^+ &\rightarrow& \sqrt{3 \over 2}\cdot{1 \over R} 
E\left( {2\A^2 \over 3}\ln {\A^2 \over gR^2} \right)
\sim  -\sqrt{3 \over 2}\cdot {\pi \over 3R} 
\left( {2\A^2 \over 3}\ln {\A^2 \over gR^2} 
\right)^{3 \over 2} \nn
V_A^- &\rightarrow& {1 \over 4\pi \sqrt 2}\cdot 
{f_T(1,1) \over R }
\eea
when $R\rightarrow 0$.
Therefore the asymptotic behavior 
of the total one-loop potential 
\be
\label{Veff}
V_T= 2k\sqrt{R^4 + {\A^4 \over 2g^2}} + V_X + V_A
\ee
is dominated by the classical 
term (the first term in (\ref{Veff}) 
when $R\rightarrow \infty$:
\be
\label{Rinf}
V \rightarrow 2kR^2 
\ee
and dominated by $V_A^+$
\be
\label{R0}
V \rightarrow  -\sqrt{3 \over 2}\cdot{\pi \over 3R} 
\left( {2\A^2 \over 3}\ln {\A^2 \over gR^2} 
\right)^{3 \over 2}
\ee 
at small R.
This tells the one-loop potential is unstable near $R\sim 0$.
Not exact but rough behavior of the potential is given 
in Fig.1.

We may also consider the static potential 
in the large $d$ approximation
 as in Ref.\cite{10}. It is difficult, 
however, to formulate it consistently with 
the definition of $\str$.

In summary, we have shown that one-loop potential 
for SU(2) non-abelian toroidal D-brane is not stable.
It is in close analogy with instability of chromomagnetic vacuum in 
SU(2) gauge theory \cite{13}.
It is much likely that in order to get the stability 
(again like in case of usual gauge theory \cite{14})
one has to consider another gauge group and (or) another 
background field.

\noindent
{\bf Acknowledgments} We thank A. Sugamoto for useful remarks, 
 A.Tseytlin for valuable comments and referee for pointing out 
mistake in original version of this work.
We are also grateful to T. Muta and whole Particle Physics Group 
at Hiroshima University for kind hospitality during the 
completing of this work.

\end{document}